\documentclass[ journal,letterpaper ]{IEEEtran}

\usepackage{graphicx,cite,epsfig,amssymb,amsmath,multicol,subfigure,mathtools,bm,mathrsfs,setspace}

\allowdisplaybreaks

\begin{document}
\title{Layered BPSK for High Data Rates}
\author{
	Bingli~Jiao,
	Yuli~Yang
	and~Mingxi~Yin
	\thanks{B. Jiao ({\em corresponding author}) and M. Yin are with the School of Electronics Engineering and Computer Science, Peking University, Beijing 100871, China (email: \{ jiaobl, yinmx\}@pku.edu.cn).}
	\thanks{Y. Yang is with the Department of Electronic and Electrical Engineering, University of Chester, Chester CH2 4NU, U.K. (e-mail: y.yang@chester.ac.uk).}
	
}

\maketitle

\begin{abstract}
This paper proposes a novel transmission strategy, referred to as layered BPSK, which allows two independent symbol streams to be layered at the transmitter on a non-orthogonal basis and isolated from each other at the receiver without inter-stream interference, aiming to achieve high data rates.  To evaluate the performance of the proposed scheme, its data rate is formulated over additive white Gaussian noise (AWGN) channels.  Based on the theoretical analysis, numerical results are provided for the performance comparisons between the proposed scheme and conventional transmission schemes, which substantiate the validity of the proposed scheme.
\end{abstract}

\begin{IEEEkeywords}
Data rate, mutual information, BPSK, AWGN channel.
\end{IEEEkeywords}

\IEEEpeerreviewmaketitle

\section{Introduction}

A general description of the transmission over memoryless AWGN channels is given by  
\begin{eqnarray}
\begin{array}{l}\label{equ1}
y = x + n,
\end{array}
\end{eqnarray}
where $y$ is the received signal and $x$ is the transmitted signal.  The received AWGN component $n$ is a random variable from a normally distributed ensemble of power $\sigma_N^2$, denoted by $n \sim \mathcal{N}(0,\sigma_N^2)$.

The data rate of transmission in (\ref{equ1}) is characterized by the mutual information between $x$ and $y$, expressed as~\cite{Shannon1948}
\begin{equation}
R = {\rm{I}}(X;Y) = {\rm{H}}(Y) - {\rm{H}}(N),
\end{equation}
where ${\rm{H}}(Y)$ is the entropy of the received signal and ${\rm{H}}(N) = {\log _2} (\sqrt{2 \pi e \sigma_N^2})$ is the entropy of the AWGN.

When the transmitted signals form a Gaussian ensemble of power $\sigma_X^2$, i.e., $x \sim \mathcal{N}(0,\sigma_X^2)$, the AWGN channel capacity, i.e., the maximum data rate of the transmission over AWGN channels, is achieved at
\begin{equation}
C = \log_2 (1 + \rho) = \log_2 (1 + {\sigma_X^2}/{\sigma_N^2})
\end{equation}
in the unit of [bits/sec/Hz], where the signal-to-noise power ratio (SNR) $\rho = {\sigma_X^2}/{\sigma_N^2}$.

\begin{figure}[!t]
	\centering
	\includegraphics[width=0.5\textwidth]{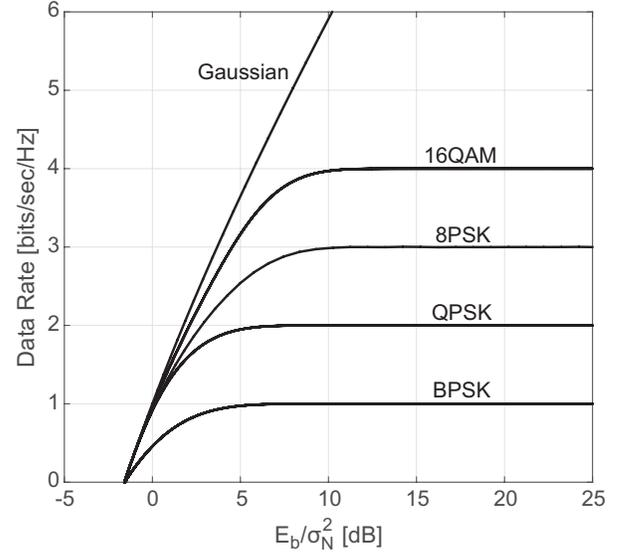}
	\caption{Data rates achieved by the Gaussian-distributed input and 16QAM-, 8PSK-, QPSK- and BPSK-modulated signals over AWGN channels versus $E_b/\sigma_N^2$.}
	\label{fig1}
\end{figure}

In Fig.~\ref{fig1}, the data rates achieved by the Gaussian-distributed input and finite-alphabet inputs, e.g., 16QAM-, 8PSK-, QPSK- and BPSK-modulated signals, over AWGN channels are plotted as the functions of $E_b/\sigma_N^2$ in [dB], where $E_b$ denotes the energy per bit.  With the Gaussian-distributed input, $E_b/\sigma_N^2 = \rho/C = \rho/\log_2 (1 + \rho)$. As $C$ goes to zero, i.e., $\rho$ goes to zero as well, $E_b/\sigma_N^2 \doteq -1.59$dB, which is the fundamental limit to reliably transmit one bit of information per unit resource, i.e., the absolute minimum ratio of bit energy to noise power required for reliable transmission of 1 bit/sec/Hz.

As shown in Fig.~\ref{fig1}, the data rates achieved by finite-alphabet inputs approach the AWGN channel capacity, i.e., the case of the Gaussian-distributed input, when $E_b/\sigma_N^2$ goes to the fundamental limit.  The main reason behind this is an important feature of the logarithm, i.e., when the SNR $\rho$ is small, the logarithmic function $\log_2 (1+\rho)$ approximates to a linear function of $\rho$~\cite{math}.  Since the data rates of finite-alphabet inputs can be expressed by logarithmic functions of their received SNRs as well, they will access the receiver at a similar data rate with the Gaussian-distributed input when the received SNRs are small.

Another important feature of the logarithm is its strict concavity.  If $\rho = \rho_1 + \rho_2$, we have $\log_2(\rho) \leq \log_2(\rho_1) + \log_2(\rho_2)$.  As such, the transmission of the information source in multiple data streams can be exploited to achieve higher data rate than that of a single data stream.  To this extent, the data rates achieved by finite-alphabet inputs are expected to beat the channel capacity~\cite{jiao}.
Over the past decades, orthogonal codes are used to combine multiple data streams over robust transmissions against interference~\cite{ortho1, ortho2}.  However, in this way, the overall data rate is not increased at all, because the orthogonality in Euclidean space isolates the transmitted power of each data stream at the receiver.

The relaxation of non-orthogonal combination provides more degrees of freedom for the system design to improve data rate through simultaneous transmission of multiple non-orthogonal data steams, where the inter-stream interference is the main challenge facing decoding performance and, meanwhile, brings an opportunity to enable the transmitted power shared among multiple data streams.  Motivated by this, in this paper we embed interference alignment into weighted sum, proposing a novel transmission strategy, referred to as layered BPSK, which allows the layering of two independent data streams on a non-orthogonal basis and demodulating with the transmitted power sharing but without the inter-stream interference.

In the following, we will detail the proposed scheme and its performance in the metric of achievable data rate.  Throughout this paper, the following mathematical notations are used: Vectors are denoted by boldface lowercase letters.  The real part and the imaginary part of a complex number are denoted by $\textrm{real}(\cdot)$ and $\textrm{imag}(\cdot)$, respectively. Moreover, $p(\cdot)$ denotes the probability density function (PDF) of a random variable, and $\mathcal{E} \{ \cdot \}$ represents the expectation (mean) operator. 


\section{Layered BPSK}

In this section, the transmission strategy referred to as layered BPSK is proposed to transmit the information source with the layering of two independent data streams for the purpose of achieving high data rate.  Herein, the basic strategy of the proposed scheme is presented firstly as one-dimensional layered BPSK and, then, it will be generalized into two-dimensional layered BPSK to double the achievable data rate.

\subsection{One-Dimensional Layered BPSK}
\label{secTx}

Consider a memoryless information source modulated with BPSK and conveyed over an AWGN channel.  The modulated BPSK symbols are divided into two independent symbol streams, denoted by $\textbf{x} = [x_1, x_2, \cdots]$ and $\textbf{z} = [z_1, z_2, \cdots]$, where the BPSK symbols $x_k, z_k \in \{+1, -1\}$, $k = 1, 2, \cdots$.

In the proposed scheme, the transmitter layers $\textbf{x}$ and $\textbf{z}$ with various weights.  Thus, the transmitted signal is a weighted sum of a symbol in $\textbf{x}$ and a symbol in $\textbf{z}$, where the two weights are contained by a $1 \times 2$ row vector $\textbf{w}$.  In particular, each symbol in the stream $\textbf{z}$ will be transmitted twice in adjacent symbol periods.

Without loss of generality, the transmissions in the $(2m-1)^\textrm{th}$ and the $(2m)^\textrm{th}$ symbol periods will be detailed to illustrate the proposed transmission strategy, $m = 1, 2, \cdots$.

In the $(2m-1)^\textrm{th}$ symbol period, the weighted sum of the $(2m-1)^\textrm{th}$ symbol in $\textbf{x}$, i.e., $x_{2m-1}$, and the $m^\textrm{th}$ symbol in $\textbf{z}$, i.e., $z_m$, is transmitted.  Therefore, the received signal is
\begin{eqnarray}
\begin{array}{l}\label{eqT1}
y_{2m-1}=\textbf{w} \begin{bmatrix}
           x_{2m-1} \\
           z_m
         \end{bmatrix} + n_{2m-1},
\end{array}
\end{eqnarray}
where $y_{2m-1}$ and $n_{2m-1} \sim \mathcal{N}(0,\sigma_N^2)$ are the received signal and AWGN component, respectively, in the $(2m-1)^\textrm{th}$ symbol period.

In the $(2m)^\textrm{th}$ symbol period,  the weighted sum of the $(2m)^\textrm{th}$ symbol in $\textbf{x}$, i.e., $x_{2m}$, and $z_m$ is transmitted, and the received signal is given by
\begin{eqnarray}
\begin{array}{l}\label{eqT2}
y_{2m}= \textbf{w} \begin{bmatrix}
           x_{2m} \\
           z_m
         \end{bmatrix} + z_m + n_{2m},
\end{array}
\end{eqnarray}
where $y_{2m}$ and $n_{2m} \sim \mathcal{N}(0,\sigma_N^2)$ are the received signal and AWGN component, respectively, in the $(2m)^\textrm{th}$ symbol period.

The key to the proposed layered BPSK lies in the design of the weights.  The $1 \times 2$ row vector consisting of the two weights is defined by
\begin{equation}
\textbf{w} = \left\{ \,
\begin{IEEEeqnarraybox}[][c]{s?s}
\IEEEstrut
$[\alpha, \beta]$ & if $x_{2m-1} \neq x_{2m}$, \\
$[\alpha, 0]$ & if $x_{2m-1} = x_{2m} = z_m$, \\
$[0, \beta/2]$ & if $x_{2m-1} = x_{2m} \neq z_m$,
\IEEEstrut
\end{IEEEeqnarraybox} 
\right.
\end{equation}
where $\alpha$ and $\beta$ are two positive real numbers ($\alpha > \beta >0$).  In detail, Table I lists the 8 equiprobable cases of the transmitted signals in the $(2m-1)^\textrm{th}$ and the $(2m)^\textrm{th}$ symbol periods, $m = 1, 2, \cdots$.

\newcommand{\tabincell}[2]{\begin{tabular}{@{}#1@{}}#2\end{tabular}}
\begin{table}[htb]
	\renewcommand{\arraystretch}{1.5}
	\centering
	\small
	\caption{Equiprobable Transmitted Signals in Layered BPSK}
	\label{Table1}
	\begin{tabular}{c | c c c | c c}
		\hline
                     & & & & & \\ [-1.2em]
		\tabincell{c}{Case} & \tabincell{c}{$x_{2m-1}$} & \tabincell{c}{$x_{2m}$} & \tabincell{c}{$z_m$} & \tabincell{c}{$\displaystyle{\renewcommand{\arraystretch}{1} \textbf{w}\begin{bmatrix} x_{2m-1} \\ z_m \end{bmatrix}}$} & \tabincell{c}{$\displaystyle{\renewcommand{\arraystretch}{1} \textbf{w}\begin{bmatrix} x_{2m} \\ z_m \end{bmatrix}}$} \\
                     & & & & & \\ [-1.2em]
            	\hline
		\tabincell{c}{1} & \tabincell{c}{$+1$} & \tabincell{c}{$-1$} & \tabincell{c}{$+1$} & \tabincell{c}{$\alpha+\beta$} & \tabincell{c}{$-\alpha+\beta$} \\
		\hline
		\tabincell{c}{2} & \tabincell{c}{$+1$} & \tabincell{c}{$-1$} &\tabincell{c}{ $-1$} & \tabincell{c}{$\alpha-\beta$} & \tabincell{c}{$-\alpha-\beta$} \\
		\hline
		\tabincell{c}{3} & \tabincell{c}{$-1$} & \tabincell{c}{$+1$} & \tabincell{c}{$+1$} & \tabincell{c}{$-\alpha+\beta$} & \tabincell{c}{$\alpha+\beta$} \\
		\hline
		\tabincell{c}{4} & \tabincell{c}{$-1$} & \tabincell{c}{$+1$} & \tabincell{c}{$-1$} & \tabincell{c}{$-\alpha-\beta$} & \tabincell{c}{$\alpha-\beta$} \\
		\hline
		\tabincell{c}{5} & \tabincell{c}{$+1$} & \tabincell{c}{$+1$} & \tabincell{c}{$+1$} & \tabincell{c}{$\alpha$} & \tabincell{c}{$\alpha$} \\
		\hline
		\tabincell{c}{6} & \tabincell{c}{$-1$} & \tabincell{c}{$-1$} & \tabincell{c}{$-1$} & \tabincell{c}{$-\alpha$} & \tabincell{c}{$-\alpha$} \\
		\hline
		\tabincell{c}{7} & \tabincell{c}{$+1$} & \tabincell{c}{$+1$} & \tabincell{c}{$-1$} & \tabincell{c}{$-\beta/2$} & \tabincell{c}{$-\beta/2$} \\
		\hline
		\tabincell{c}{8} & \tabincell{c}{$-1$} & \tabincell{c}{$-1$} & \tabincell{c}{$+1$} & \tabincell{c}{$\beta/2$} & \tabincell{c}{$\beta/2$} \\
		\hline
	\end{tabular}
\end{table}

Upon receiving $y_{2m-1}$ and $y_{2m}$ in the $(2m-1)^\textrm{th}$ and $(2m)^\textrm{th}$ symbol periods, the receiver will firstly demodulate $z_m$ according to
\begin{equation} \label{eqRxS}
\tilde{z}_m = y_{2m-1} + y_{2m}.
\end{equation}
This demodulation is exactly the same as the demodulation of conventional BPSK: If $\tilde{z}_m > 0$, we choose $\hat{z}_m = +1$; otherwise, $\hat{z}_m = -1$.

Then, the symbols $x_{2m-1}$ and $x_{2m}$ will be demodulated by a conventional BPSK demodulator as well, based on the demodulated symbol $\hat{z}_m$.  To demodulate $x_{2m-1}$, the signal sent to the BPSK demodulator is
\begin{equation} \label{eqRxRm1}
\tilde{x}_{2m-1} = y_{2m-1} - \hat{z}_m \beta.
\end{equation}
As a result, we have $\hat{x}_{2m-1} = +1$ if $\tilde{x}_{2m-1}>0$ or $\hat{x}_{2m-1} = -1$ if $\tilde{x}_{2m-1}<0$.

Similarly, $x_{2m}$ will be decoded by the conventional BPSK demodulator via the signal given by
\begin{equation} \label{eqRxRm2}
\tilde{x}_{2m} = y_{2m} - \hat{z}_m \beta.
\end{equation}
Subsequently, $\hat{x}_{2m} = +1$ if $\tilde{x}_{2m}>0$ or $\hat{x}_{2m} = -1$ if $\tilde{x}_{2m}<0$.

\subsection{Two-Dimensional Layered BPSK}

Now, we generalize the basic one-dimensional layered BPSK into the two-dimensional scenario, by introducing another dimension composed of BPSK symbols $\{+j,-j\}$, where $j = \sqrt{-1}$ stands for the imaginary unit.  In this dimension, another memoryless information source is modulated with $\{+j,-j\}$, and the modulated BPSK symbols are divided into two independent symbol streams as well, denoted by $\textbf{x}' = [x'_1, x'_2, \cdots]$ and $\textbf{z}'=[z'_1, z'_2, \cdots]$, where $x'_k \in \{+j, -j\}$ and $z'_k \in \{+j, -j\}$, $k = 1, 2, \cdots$.

Moreover, $\textbf{x}'$ and $\textbf{z}'$ are layered in the same way as $\textbf{x}$ and $\textbf{z}$ are layered, with various weights.  The two weights in the layering of $\textbf{x}'$ and $\textbf{z}'$ are contained by a $1 \times 2$ row vector $\textbf{w}'$.

As the dimension composed of $\{+j,-j\}$ and the dimension composed of $\{+1, -1\}$ are orthogonal, the layered $\textbf{x}'$ and $\textbf{z}'$ will be transmitted together with the layered $\textbf{x}$ and $\textbf{z}$.

With the two-dimensional layered BPSK, the received signal in the $(2m-1)^\textrm{th}$ symbol period is expressed as
\begin{eqnarray}
\begin{array}{l}\label{T2m1}
y'_{2m-1}=\textbf{w} \begin{bmatrix}
           x_{2m-1} \\
           z_m
         \end{bmatrix} + \textbf{w}' \begin{bmatrix}
           x'_{2m-1} \\
           z'_m
         \end{bmatrix} + n'_{2m-1},
\end{array}
\end{eqnarray}
where $y'_{2m-1}$ and $n'_{2m-1} \sim \mathcal{CN}(0,\sigma_N^2)$ are the received signal of the two-dimensional layered BPSK and the observed complex AWGN component, respectively, in the $(2m-1)^\textrm{th}$ symbol period.

In the $(2m)^\textrm{th}$ symbol period, the received signal is given by
\begin{eqnarray}
\begin{array}{l}\label{T2m2}
y'_{2m}=\textbf{w} \begin{bmatrix}
           x_{2m} \\
           z_m
         \end{bmatrix} + \textbf{w}' \begin{bmatrix}
           x'_{2m} \\
           z'_m
         \end{bmatrix} + n'_{2m},
\end{array}
\end{eqnarray}
where $y'_{2m}$ and $n'_{2m} \sim \mathcal{CN}(0,\sigma_N^2)$ are the received signal of the two-dimensional layered BPSK and the observed complex AWGN component, respectively, in the $(2m)^\textrm{th}$ symbol period.

In (\ref{T2m1}) and (\ref{T2m2}), the $1 \times 2$ row vector $\textbf{w}'$, consisting of the two weights in the layering of $\textbf{x}'$ and $\textbf{z}'$, is designed as
\begin{equation}
\textbf{w}' = \left\{ \,
\begin{IEEEeqnarraybox}[][c]{s?s}
\IEEEstrut
$[\alpha', \beta']$ & if $x'_{2m-1} \neq x'_{2m}$, \\
$[\alpha', 0]$ & if $x'_{2m-1} = x'_{2m} = z'_m$, \\
$[0, \beta'/2]$ & if $x'_{2m-1} = x'_{2m} \neq z'_m$,
\IEEEstrut
\end{IEEEeqnarraybox} 
\right.
\end{equation}
where $\alpha'$ and $\beta'$ are two positive real numbers ($\alpha' > \beta' >0$).  

Subsequently, the receiver will get the signal $\tilde{z}'_m = y'_{2m-1} + y'_{2m}$ to demodulate $z_m$ according to the real part of $\tilde{z}'_m$ and $z'_m$ according to the imaginary part of $\tilde{z}'_m$.  As a result, we have $\hat{z}_m = +1$ if $\textrm{real}(\tilde{z}'_m) > 0$, or $\hat{z}_m = -1$ if $\textrm{real}(\tilde{z}'_m) < 0$.  Meanwhile, we have $\hat{z}'_m = +1$ if $\textrm{imag}(\tilde{z}'_m) > 0$, or $\hat{z}'_m = -1$ if $\textrm{imag}(\tilde{z}'_m) < 0$.

By generating the signals $\tilde{x}'_{2m-1} = y'_{2m-1} - (\hat{z}_m \beta + \hat{z}'_m \beta')$ and $\tilde{x}'_{2m} = y'_{2m} - (\hat{z}_m \beta + \hat{z}'_m \beta')$, the BPSK symbols $x_{2m-1}$, $x_{2m}$, $x'_{2m-1}$, $x'_{2m}$ will be demodulated as the following:
\begin{equation}
\hat{x}_{2m-1} = \left\{ \,
\begin{IEEEeqnarraybox}[][c]{s?s}
\IEEEstrut
$+1$ & if $\textrm{real}(\tilde{x}'_{2m-1})>0$, \\
$-1$ & if  $\textrm{real}(\tilde{x}'_{2m-1})<0$;
\IEEEstrut
\end{IEEEeqnarraybox} 
\right.
\end{equation}

\begin{equation}
\hat{x}_{2m} = \left\{ \,
\begin{IEEEeqnarraybox}[][c]{s?s}
\IEEEstrut
$+1$ & if $\textrm{real}(\tilde{x}'_{2m})>0$, \\
$-1$ & if  $\textrm{real}(\tilde{x}'_{2m})<0$;
\IEEEstrut
\end{IEEEeqnarraybox} 
\right.
\end{equation}

\begin{equation}
\hat{x}'_{2m-1} = \left\{ \,
\begin{IEEEeqnarraybox}[][c]{s?s}
\IEEEstrut
$+1$ & if $\textrm{imag}(\tilde{x}'_{2m-1})>0$, \\
$-1$ & if  $\textrm{imag}(\tilde{x}'_{2m-1})<0$;
\IEEEstrut
\end{IEEEeqnarraybox} 
\right.
\end{equation}

\begin{equation}
\hat{x}'_{2m} = \left\{ \,
\begin{IEEEeqnarraybox}[][c]{s?s}
\IEEEstrut
$+1$ & if $\textrm{imag}(\tilde{x}'_{2m})>0$, \\
$-1$ & if $\textrm{imag}(\tilde{x}'_{2m})<0$.
\IEEEstrut
\end{IEEEeqnarraybox} 
\right.
\end{equation}


\section{Achievable Data Rates}

To evaluate the performance of the proposed layered BPSK, the analysis of its data rate over AWGN channels is established in this section.

For a one-dimensional layered BPSK system, there are two input spaces, $X$ and $Z$, and one output space $Y$.  Grounded on the definition of mutual information between the input and output of a system, the data rate of one-dimensional layered BPSK, denoted by $R_1$, is formulated by
\begin{equation} \label{eqR1}
R_1 = {\rm{I}}(X, Z;Y) =  {\rm{I}}(X;Y|Z) + {\rm{I}}(Z;Y).
\end{equation}

In the BPSK demodulation of the input space $X$ given by~(\ref{eqRxRm1}) and (\ref{eqRxRm2}), the signals sent to the demodulator are the sum of the AWGN and 4 equiprobable BPSK signals with various amplitudes, which are deemed to be 4 random variables with different mean values, denoted by $\zeta_i$, $i = 1, 2, 3, 4$, and their PDFs are
\begin{subequations}
\begin{alignat}{4}
p(\zeta_1) &= \frac{1}{2} \frac{1}{\sqrt{2 \pi \sigma_N^2}}\left( e^{-\frac{(\zeta_1-\alpha)^2}{2 \sigma_N^2}} + e^{-\frac{(\zeta_1+\alpha)^2}{2 \sigma_N^2}}\right) , \\
p(\zeta_2) &= \frac{1}{2} \frac{1}{\sqrt{2 \pi \sigma_N^2}}\left( e^{-\frac{(\zeta_2-\alpha)^2}{2 \sigma_N^2}} + e^{-\frac{(\zeta_2+\alpha)^2}{2 \sigma_N^2}}\right) , \\
p(\zeta_3) &= \frac{1}{2} \frac{1}{\sqrt{2 \pi \sigma_N^2}}\left( e^{-\frac{(\zeta_3-\alpha+\beta)^2}{2 \sigma_N^2}} + e^{-\frac{(\zeta_3+\alpha-\beta)^2}{2 \sigma_N^2}}\right) , \\
p(\zeta_4) &= \frac{1}{2} \frac{1}{\sqrt{2 \pi \sigma_N^2}}\left( e^{-\frac{(\zeta_4-\beta/2)^2}{2 \sigma_N^2}} + e^{-\frac{(\zeta_4+\beta/2)^2}{2 \sigma_N^2}}\right).
\end{alignat}
\end{subequations}
Thus, the item ${\rm{I}}(X;Y|Z)$ in (\ref{eqR1}) can be expressed by
\begin{equation}
{\rm{I}}(X;Y|Z) = \frac{1}{4} \sum_{i = 1}^4 \mathcal{E} \{- \log_2 p(\zeta_i) \} - {\log _2} (\sqrt{2 \pi e \sigma_N^2}),
\end{equation}
where the first item in the right-hand side is the entropy of the signal space formed by the signals sent to the BPSK demodulator given in (\ref{eqRxRm1}) and (\ref{eqRxRm2}).  In detail, the expectation $\mathcal{E} \{- \log_2 p(\zeta_i) \} = -\int_{-\infty}^{\infty} p(\zeta_i)\log_2 p(\zeta_i)\textrm{d}\zeta_i$, $i = 1, 2, 3, 4$.

As each signal in the input space $Z$ is transmitted twice, the item ${\rm{I}}(Z;Y)$ in (\ref{eqR1}) can be calculated, on the basis of the BPSK demodulation in (\ref{eqRxS}), as 
\begin{equation}
{\rm{I}}(Z;Y) = \frac{1}{2} \left(\frac{1}{4} \sum_{i = 1}^4 \mathcal{E} \{- \log_2 p(\xi_i) \} - {\log _2} (\sqrt{4 \pi e \sigma_N^2})\right),
\end{equation}
where the item $(1/4) \sum_{i = 1}^4 \mathcal{E} \{-\log_2 p(\xi_i) \}$ is the entropy of the signal space formed by $\tilde{z}_m$ given in (\ref{eqRxS}), $m = 1, 2, \cdots$, and the expectation $\mathcal{E} \{-\log_2 p(\xi_i) \} = -\int_{-\infty}^{\infty} p(\xi_i)\log_2 p(\xi_i) \textrm{d}\xi_i$.  The variables $\xi_i$, $i = 1, 2, 3, 4$, are the signals sent to the BPSK demodulator (\ref{eqRxS}) with different mean values and their PDFs are given by
\begin{subequations}
\begin{alignat}{4}
p(\xi_1) &= \frac{1}{2} \frac{1}{\sqrt{4 \pi \sigma_N^2}}\left( e^{-\frac{(\xi_1-2\beta)^2}{4 \sigma_N^2}} + e^{-\frac{(\xi_1+2\beta)^2}{4 \sigma_N^2}}\right) , \\
p(\xi_2) &= \frac{1}{2} \frac{1}{\sqrt{4 \pi \sigma_N^2}}\left( e^{-\frac{(\xi_2-2\beta)^2}{4 \sigma_N^2}} + e^{-\frac{(\xi_2+2\beta)^2}{4 \sigma_N^2}}\right) , \\
p(\xi_3) &= \frac{1}{2} \frac{1}{\sqrt{4 \pi \sigma_N^2}}\left( e^{-\frac{(\xi_3-2\alpha)^2}{4 \sigma_N^2}} + e^{-\frac{(\xi_3+2\alpha)^2}{4 \sigma_N^2}}\right) , \\
p(\xi_4) &= \frac{1}{2} \frac{1}{\sqrt{4 \pi \sigma_N^2}}\left( e^{-\frac{(\xi_4-\beta)^2}{4 \sigma_N^2}} + e^{-\frac{(\xi_4+\beta)^2}{4 \sigma_N^2}}\right).
\end{alignat}
\end{subequations}

In the case of two-dimensional layered BPSK, the data rate of layered $\textbf{x}'$ and $\textbf{z}'$ accessing the receiver is denoted by $R'_1$, which can be obtained in the same way as the formulation of $R_1$.  As a consequence, the data rate of two-dimensional layered BPSK, denoted by $R_2$, is achieved at
\begin{equation}
R_2 = R_1 + R'_1,
\end{equation}
where $R_1$ is given by (\ref{eqR1}) and $R'_1$ is obtained by simply changing $\alpha$ and $\beta$ in the calculation of $R_1$ with $\alpha'$ and $\beta'$, respectively.  If $\alpha=\alpha'$ and $\beta = \beta'$, we will have $R_2 = 2 R_1$.


\section{Numerical Results and Discussions}

Herein, we numerically illuminate the performance of the proposed layered BPSK in the metric of data rate and compare it with conventional modulation schemes.

\begin{figure}[!t]
	\centering
	\includegraphics[width=0.5\textwidth]{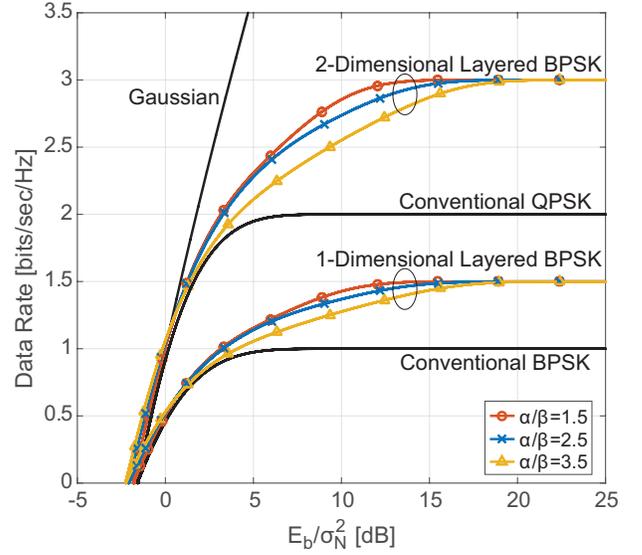}
	\caption{Data rate comparisons between the proposed layered BPSK and conventional transmission schemes.}
	\label{fig2}
\end{figure}

In Fig.~\ref{fig2}, the data rates of layered BPSK with various ratios of $\alpha/\beta$ are compared with conventional BPSK and QPSK schemes over AWGN channels, where $\alpha = \alpha'$ and $\beta = \beta'$.  As expected, the data rate of one-dimensional layered BPSK converges to 1.5 bits/sec Hz, and the data rate of two-dimensional layered BPSK converges to 3 bits/sec/Hz.  Specifically around the fundamental limit, the data rate of the layered BPSK is slightly higher than the AWGN channel capacity.  Moreover, with the increase in the ratio of $\alpha/\beta$, the data rate of the layered BPSK increases at low SNRs but decreases at high SNRs.

At low SNRs, the competitive advantage of the layered BPSK comes from the cross utilization of transmitted power in the BPSK demodulation of either layer.  Elaborating slightly further, the equivalent SNRs in the BPSK demodulation of each $x_k$ and each $z_k$, $k = 1, 2, \cdots$, are
\begin{equation}
\rho_x = \frac{1}{\sigma_N^2} \left[\frac{1}{2}\alpha^2 + \frac{1}{4}(\alpha-\beta)^2+\frac{1}{4}(\frac{\beta}{2})^2\right]
\end{equation}
and
\begin{equation}
\rho_z = \frac{1}{2 \sigma_N^2} \left[\frac{1}{2}(2\beta)^2 + \frac{1}{4}(2\alpha)^2+ \frac{1}{4}\beta^2 \right],
\end{equation}
respectively.  On the other hand, the average received SNR of the layered BPSK is constrained by the received SNR in the conventional BPSK demodulation, denoted by $\rho_\textrm{BPSK}$, i.e.,
\begin{equation}
\rho_\textrm{BPSK} \triangleq \frac{1}{\sigma_N^2} \left[\frac{1}{2}(\alpha^2 + \beta^2) + \frac{1}{4}\alpha^2+ \frac{1}{4}(\frac{\beta}{2})^2 \right].
\end{equation}
Evidently, $\rho_x + (1/2)\rho_z > \rho_\textrm{BPSK}$.  As shown in Fig.~\ref{fig1}, in the low-SNR region, the data rates of finite-alphabet inputs approximate to linear functions of their SNRs.  Therefore, the received SNRs can be used to scale the data rates in the low-SNR region.  As a consequence, the one-dimensional layered BPSK achieves higher data rate than conventional BPSK.  Since $R_2 = 2 R_1$ herein and the data rate of QPSK is twice that of BPSK, the two-dimensional layered BPSK achieves higher data rate than conventional QPSK.

\section{Conclusion}
In this paper, the layered BPSK was proposed to layer two independent symbol streams at the transmitter on a non-orthogonal basis and demodulate the layering without inter-stream interference at the receiver.  For the purpose of demonstrating the advantages of the proposed scheme, its data rate was established to facilitate the performance evaluation.  Numerical results on the performance comparisons between the proposed scheme and conventional transmission schemes not only substantiated the validity of the proposed scheme, but also provided useful reference for the power allocation of the proposed scheme.  Moreover, it is worth mentioning that the data rate of the layered BPSK is slightly higher than the AWGN channel capacity around the fundamental limit.

\end{document}